\documentclass[aps,reprint,prl,groupedaddress,longbibliography]{revtex4-1}
\usepackage[utf8]{inputenc}
\usepackage{amsmath,amssymb,mathrsfs,amsthm}
\usepackage{bbm}
\usepackage{chemarrow}
\usepackage{graphicx}
\usepackage[colorlinks=true,allcolors=blue]{hyperref}
\usepackage{soul}
\def\rd{{\rm d}}

\def\vp{{\bf p}}

\def\vx{{\bf x}}
\def\vy{{\bf y}}

\def\vmu{\boldsymbol{\mu}}
\def\vnu{\boldsymbol{\nu}}

\def\vdelta{\boldsymbol{\delta}}

\def\vpi{\boldsymbol{\pi}}

\def\mG{{\bf G}}

\begin{document}

\title{Internal Energy, Fundamental Thermodynamic Relation, and\\ Gibbs' Ensemble Theory as Laws of Statistical Counting}
\author{Hong Qian}
\email{hqian@uw.edu}
\affiliation{Department of Applied Mathematics, University of Washington, Seattle, WA 98195-3925, USA}

\date{\today}

\begin{abstract}
Counting {\it ad infinitum} is the holographic observable to a statistical dynamics with finite states under independent repeated sampling.  Entropy provides the infinitesimal probability for an observed frequency $\hat{\vnu}$ w.r.t. a probability prior $\vp$. Following Callen's postulate and through Legendre-Fenchel transform, without help from mechanics, we show an internal energy $\vmu$ emerges; it provides a linear representation of real-valued observables with full or partial information. Gibbs' fundamental thermodynamic relation and theory of ensembles follow mathematically. $\vmu$ is to $\hat{\vnu}$ what $\omega$ is to $t$ in Fourier analysis.  

\end{abstract}

\maketitle



Sometime a mathematical transform can provide a fundamental concept beyond just being a technique for solving a problem, and through which a understanding of natural phenomena emerges.  A case in point is the Fourier transform (FT) that leads to the theory of harmonics in music instruments \cite{music-siam} and the very concept of optical spectrum.  FT represents a function of time $f(t)$ in terms of $\tilde{f}(\omega)$, where $\omega$ is introduced as a novel notion, the frequency of a sinusoidal oscillatory component in time \cite{fourier-book}.  The solutions to a large class of problems in differential calculus involving $t$ can be very efficiently expressed through FT.

We show here that the notion of {\em internal energy} first appeared in the theory of thermodynamics in the $19^{th}$ century, collectively developed by J. R. von Mayer, W. Rankine, R. Clausius, and W. Thomson among many others \cite{truesdell-book}, is a concept that can be understood, and generalized, in statistical counting. The transformation in question is the Legendre-Fenchel transform (LFT) \cite{lu-qian-22,bedeaux}, a more refined mathematical formulation of the traditional Legendre transform \cite{rockafellar-book}. 

When a simple statistical analysis is carried out on a set of data, correlated or not, it is usually assumed that they are from an identical probability distribution. One of the best understood systems that exhibit an invariant probability is an {\em ergodic dynamical system} \cite{qianmin-book}.  The ergodic theory of classical Hamiltonian dynamics has been an intense research area in both physics and mathematics for more than a century \cite{dorfman-book,mackey}. Even when the data are from seemingly different ``subjects'', say different individuals within a biological species, it is understood that an ergodic mating or mutational process is behind the statistical practice; and the conclusions drawn are most meaningful in this regard. Such an ergodic stochastic dynamic perspective has transformed cell biology in recent years \cite{QianGeBook}. 

Let us consider the repeated statistical samples {\em ad infinitum} of a system with finite state space $\mathcal{S}=\{0,1,\cdots,n\}$. In the present work we shall restrict our discussion for independent and identically distributed (i.i.d.) samples.  More general sampling of Markov data will be published elsewhere.  The number counting $\vnu=(\nu_0,\cdots,\nu_n)$ with $\nu_0+\cdots+\nu_n=N$ and counting frequency $\hat{\vnu}=\vnu/N$, not to be confused with the $\omega$ in FT above, has a homogeneous degree $1$ neg-entropy function with respect to a given probability prior $\vp=(p_0,\cdots,p_n)$ 
\cite{dembo-book,JaynesBook}:
\begin{equation}
\label{ent-func} 
 \Phi(\vnu) = \sum_{i=0}^n \nu_i
 \ln\left(\frac{\nu_i}{Np_i} \right).
\end{equation} 
The appendix provides the mathematical origin of the non-negative $\Phi(\vnu)$ as a result of statistical counting.  In information theory, it is interpreted as the ``surprise'' in observing the $\vnu$ under the assumption $\vp$
\cite{levine,bennaim-book}.  It is a double-edged sword which tells the rareness of $\vnu$ (or $\hat{\vnu}$) w.r.t. $\vp$ or erroneous of $\vp$ w.r.t. $\vnu$.  For any statistical modeling, the prior probability $\vp$ needs not to be realistic; it simply provides a starting point for analyzing data statistically; it can and should be updated when confident, meaningful observations are made on a system. The confidence usually comes from big data, {\em e.g.} a large $N$, that we assume throughout the paper.

The entropy in (\ref{ent-func}), therefore, is the fundamental statistical prior for an i.i.d. sample \cite{JaynesBook,dembo-book}; it characterizes the {\em relationship} between $\vnu$ and $\vp$ in the sampling process.  It is an Eulerian degree $1$ homogeneous function of $\vnu$: $\Phi(\lambda\vnu)=\lambda\Phi(\vnu)$.  This fits naturally to the fundamental thermodynamic postulate formulated by H. B. Callen \cite{callen-book}.  The LFT of $\Phi$ as a function of the normalized $\hat{\vnu}$ then yields \cite{lu-qian-22,cyq-21,bedeaux}:
\begin{subequations}
\label{fe-func}
\begin{equation}
 \Psi(\vmu) = \inf_{\hat{\vnu}}
 \left\{\sum_{i=0}^n \hat{\nu}_i\mu_i
   +\Phi\big(\hat{\vnu}\big) \right\}
= -\ln\sum_{i=0}^n p_ie^{-\mu_i},
\end{equation}
with corresponding optimal $\hat{\vnu}^*(\vmu)$
\begin{equation}
    \big(\hat{\vnu}^*\big)_i =
    \frac{p_ie^{-\mu_i}}{\sum_{\ell=0}p_{\ell} e^{-\mu_{\ell}} }, \text{ and } 
    \mu_i = -\left(\frac{\partial\Phi(\hat{\vnu}^*)}{\partial\hat{\nu}_i}\right).
\end{equation}
\end{subequations}
Note that the second equation in (\ref{fe-func}b) is obtain when one uses calculus to solve the infimum in (\ref{fe-func}a); this recovers the traditional Legendre transform.  Normalizing $\vnu$ to $\hat{\vnu}$ induces a gauge freedom in (\ref{fe-func}), an arbitrary additive constant to $\mu_i$. In statistical thermodynamics, the conjugate variable $\mu_k$ introduced in Eq. \ref{fe-func} has been interpreted as the {\em internal energy} of the state $k$, in $k_BT$ unit \cite{qian_jctc}; then $\hat{\vnu}\cdot\vmu$ is the mean internal energy of ``the statistical system''.

In a real-world laboratory working on a particular system, the $\vnu$'s tend to infinity as $N\to\infty$ but $\hat{\vnu}$ converges to the intrinsic property of the statistical system.  The assumed $\vp$ then is expected to be replaced by the observed, real $\hat{\vnu}$ according to Bayesian statistical logic \cite{cyq-21,qian_jctc}. This concludes the statistical investigation of the particular system w.r.t. the type of observations.  The neg-entropy function in (\ref{ent-func}) actually provides a {\em meta-statistics} for all possible observed $\hat{\vnu}$; assessing their respective infinitesimal probability (rate) w.r.t. the prior $\vp$ (see Appendix). 

Unfortunately, a complete counting for the entire state space $\mathcal{S}$ is only a {\em gedankenexperiment}.  The significance of Gibbs' ensemble theory is in dealing with observations from a small set of real-valued observables $g_1(i), g_2(i),\cdots, g_J(i)$, where $i\in\mathcal{S}$ but $J\ll n$. These $g$'s are random variables on the state space $\mathcal{S}$.  In fact, their observed mean values are linear combinations of the $\hat{\vnu}$:
\begin{equation}
\label{nu2X}
       x_j  = \sum_{i=0}^n
       \hat{\nu}_ig_j(i).
\end{equation}
To fix mathematical notations, we append $g_0(i)=1$ and  $x_0=1$, which represent the fact that $\hat{\vnu}$ is always normalized, and denote $(n+1)\times (J+1)$ matrix $\mG_J$ with elements
\begin{equation}
    (\mG_J)_{ij} = \left\{\begin{array}{ccl}
      1 && j=0,
    \\
       g_j(i) && j = 1,\cdots,J.
     \end{array} \right.
\end{equation}
Eq. \ref{nu2X} shows that if all the $g$'s are linearly independent and $J=n$, then one can solve the normalized $\hat{\vnu}$ uniquely from each set of $x$'s: $\hat{\vnu}=\vx\mG_n^{-1}$.  We refer such a set of observables {\em holographic} with full information.  In the following discussion, we shall always imagine the $(g_1,\cdots,g_J)$ as the first $J$ component of a holographic observable $(g_0,g_1,\cdots,g_n)$.  When $J<n$, there is missing information \cite{JaynesBook,qian_jctc,bennaim-book}.  

With a set of observed values $\vx'=(x_1,\cdots,x_J)$ in hand where $J<n$, the {\em maximum entropy principle} (MEP) from classical thermodynamics \cite{callen-book} and the contraction principle from the mathematical theory of probability \cite{dembo-book} assert that the most probable $\hat{\vnu}^*$ that is consistent with the set of $\vx'$ corresponds to minimum neg-entropy:  
\begin{equation}
\label{projection}
    \hat{\vnu}^* = \arg\inf_{\hat{\vnu}}\Big\{
     \Phi(\hat{\vnu})\Big|\  \hat{\vnu} 
         \mG_J = \vx' 
    \Big\}.
\end{equation}
The entire Gibbs' ensemble theory arises in solving the mathematical problem posed in Eq. (\ref{projection}) through LFT.  See Appendix for its origin.

Entropy functions for different observables are different.  First, for invertible $\mG_n$, one has the entropy function for the holographic observable $\vx=(1,x_1,\cdots,x_n)$:
\begin{equation}
    \Phi_{\vx}\big(\vx) \equiv 
    \Phi\big(\vx\mG_n^{-1}\big).
\end{equation}
This is simply a change of the independent variables from $\vnu$ to $\vx$.  Then in terms of this entropy function $\Phi_{\vx}$, (\ref{projection}) becomes 
\begin{eqnarray}
\label{eq7}
  \varphi(\vx')
    &=& \inf_{\hat{\vnu}}\Big\{\Phi\big(\hat{\vnu}\big)\Big|\ \hat{\vnu}\mG_J=\vx' \Big\}
\\
    &=& \inf_{x_{J+1},\cdots,x_n} 
    \Big\{ \Phi_{\vx}\big(\vx\big) \Big|\ x_1=x_1',\cdots,x_J=x_J'\Big\}.
\nonumber
\end{eqnarray}
Intimately related to the generating function of a probability distribution, the LFT provides a powerful mathematical transform of the entropy functions $\Phi(\vnu)$, $\Phi_{\vx}(\vx)$, and $\varphi(\vx')$ in terms of their conjugates in the energy representation: Parallel to the $\Psi(\vmu)$ in (\ref{fe-func}) are, 
\begin{eqnarray}
    \Psi_{\vy}\big(\vy) &=& 
    \inf_{\vx}\left\{ \sum_{i=1}^n
     x_iy_i + \Phi_{\vx}(\vx)
    \right\},
\label{eq8}\\    
\psi(\vy') &=& \inf_{\vx'}\left\{ 
   \sum_{j=1}^J x_j'y'_j + \varphi\big(\vx'\big)
  \right\}.
\label{phi2psi}
\end{eqnarray}
These psi's are now related through linear transformation:
\begin{equation} 
\label{eq10}
 \Psi_{\vy}(\vy)=\Psi\big(\mG_n\vy\big), 
\end{equation} 
and projection:
\begin{subequations} 
\label{eq11}
\begin{eqnarray}
 \psi(\vy') &=& \Psi_{\vy}\big(
   y_1',\cdots,y_J',0,\cdots,0\big)
   = \Psi\big(\mG_J\vy'\big) \hspace{1cm}
\\
   &=&  -\ln\sum_{i=0}^n p_i 
   \exp\left[-\sum_{j=1}g_j(i)y_j'\right].
\end{eqnarray}
\end{subequations}
And finally, since $\psi$ is convex, the inverse LFT yields
\begin{eqnarray}
    -\varphi(\vx') &=& 
      \inf_{\vy'}\left\{ \sum_{i=1}^J 
       x_i'y_i' - \psi(\vy') \right\}
\nonumber\\
    &=& \left\{\begin{array}{ccl}  
    -\varphi &=& \vy'\cdot\nabla\psi(\vy')-\psi(\vy') 
    \\[4pt]
    \vx'&=& \nabla\psi(\vy')
    \end{array}\right.
\label{psi2phi}    
\end{eqnarray}
The optimization in (\ref{projection}) is now ``solved'' completely in closed form, through LFT and its inverse, as a parametric function in terms of $\vy'$ given in (\ref{psi2phi}). 

The equation $-\varphi=\vy'\cdot\nabla\psi-\psi$ in (\ref{psi2phi}) should be recognized as a generalization of the celebrated ``entropy $=$ mean internal energy $-$ free energy'', where
\begin{equation}
    \big( \nabla\psi \big)_k = 
     \frac{ \displaystyle 
        \sum_{i=0}^n g_k(i)p_i\exp\sum_{j=1}^J
           g_j(i)y_j}{\displaystyle 
        \sum_{i=0}^n p_i\exp\sum_{j=1}^J
           g_j(i)y_j}
\end{equation}
is the mean value of $g_k$ following Eq. \ref{eq11}b, whose conjugate variable is $y'_k$.  The identification of $\vmu=\mG_J\vy'$ in (\ref{eq11}a) with the first law of thermodynamics as formulated by Gibbs seems natural.

The $y_{J+1}=\cdots=y_n=0$ in (\ref{eq11}a) has a very clear thermodynamic interpretation: Since the conjugate variable $\vy$ are the partial derivatives of the entropy function $\Phi_{\vx}$ w.r.t. $\vx$, finding $x$'s with maximum entropy in Eq. \ref{eq7} is simply setting corresponding $y=0$, {\em e.g.}, let the {\em entropic force} being zero.  For each independent observable $g_j$, $y_j$ is its ``custom-designed'' conjugate force and $y_j \times \rd x_j$ contributes a term to the internal energy as the ``thermodynamic work'' associated with $g_j$: The internal energy $\vmu$ is a highly flexible, adaptive representation of the $\hat{\vnu}$.  When $J=n$, $\vmu=\mG_n\vy$ and Eq. (\ref{eq10}) provides a complete ``detailing'' of the internal energy in terms of a set of holographic observables.
MEP is for missing information \cite{JaynesBook}.

{\bf\em Gibbs distribution and linear algebraic representation.} There is a geometric picture associated with the above ``thermodynamic analysis''. As we have stated, counting frequency {\em ad infinitum} $\hat{\vnu}$ is a fundamental, intrinsic property of an ergodic dynamical system.  The space of all possible frequency distributions $\hat{\vnu}$, with $\hat{\nu}_0+\cdots+\hat{\nu}_n=1$, is a $n$-dimensional hyper-plane in the positive quadrant of $\mathbb{R}^{n+1}$,
known as a probability simplex $\mathscr{M}_n$.  For a given set of observables $(g_1,\cdots,g_J)$, the $\mathscr{M}_n$ is foliated by $\hat{\vnu}\mG_j=\vx'$ with different $\vx'$. On each leave of the foliation there is the most probable $\vnu^*(\vx')$, which is located at the tangent point between the $(n-J)$-dimensional leave and a
$(n-1)$-dimensional level set of the $\Phi\big(\hat{\vnu}\big)$ function. At this point $\nabla_{\vnu}\Phi(\hat{\vnu})=-\vmu(\hat{\vnu})$ is the normal vector to the $\vx'$-leave in $\mathbb{R}^{n+1}$, and $\nabla\varphi(\vx')=-\vy'$ is its projection onto the $J$-manifold of $\vx'$: 
\begin{subequations}
\begin{equation}
    \vnu^*\big( \vx' \big) 
    =\left\{\begin{array}{l}
    \displaystyle \nu^*_i=  \frac{1}{Z(\vy')}p_i\exp
     \Big[-\sum_{j=1}^J g_j(i)y'_j\Big]
     \\
     \displaystyle   x'_j=
      \frac{1}{Z(\vy')}
      \sum_{i=0}^n g_j(i)p_i\exp
     \Big[-\sum_{j=1}^J g_j(i)y'_j\Big]
      \end{array}\right.
\end{equation}
in which
\begin{equation}
    Z(\vy') = \sum_{i=0}^n p_i\exp\left[
     -\sum_{i=1}^J g_j(i)y'_j
      \right].
\end{equation}
\end{subequations}
All the other points on the same $\vx'$-leave are no longer relevant: They are deemed statistically impossible under the prior $\vp$ and observed $\vx'$.  The foliation therefore represents a partition of the $\mathscr{M}_n$ into a macro- and a micro-worlds: Transversing between different $\vx'$-leaves are {\em macroscopic thermodynamic processes} that follows the $\vy'(\vx')$.  According to the logic of Bayesian statistics, one should use the most probability frequency distribution $\hat{\vnu}^*(\vx')$ to update the prior $\vp$ for the particular system with observed $\vx'$. The {\em microscopic world} is still random; due to missing information; but its prior is now updated.  This is Gibbs' statistical ensemble.

With a given set of $(g_1,\cdots,g_J)$, the $\mathscr{M}_n$ is collapsed into to $J$-manifold in $\mathbb{R}^{n+1}$, which is parametrized by the $\vx'$, or equivalently $\vy'$.  There is no uncertainty in this ``macroscopic'' description. For course for a different set of $g$'s and $J'$, there will be a different $J'$-manifold.  It will be desirable to treat different $g$'s through transformations. We note that even though $\mathscr{M}_n$ is a ``plane'' in $\mathbb{R}^{n+1}$, it is not a linear Euclidean space since for any $c\neq 1$, $c\hat{\vnu}\notin\mathscr{M}_n$.  Neither are the $\vx'$-leaves. They are affine manifolds  \cite{hqt}.  The locating of $\hat{\vnu}^*(\vx')$ is a highly nonlinear 
procedure in the space of energies.

The LFT, in terms $\Psi(\vmu)$, $\Psi_{\vy}(\vy)$ and $\psi(\vy')$ {\em etc.}, enters as a powerful algebraic linear representation of the MEP procedure.  The ``collapse'' of a holographic $\vy$ to $\vy'$ with missing information means simply neglecting all the extra dimensions: $y_{J+1}=\cdots=y_n=0$.  This is because due to the convexity of $\Phi\big(\hat{\vnu}\big)$, there is a one-to-one relation between $\hat{\vnu}$ and $\vmu=-\nabla\Phi$ under a proper gauge fixing.  And since the constrains to MEP in (\ref{projection}) are all linear due to the nature of observables being random variables, each $g$ determines a $1$-dimensional linear subspace in the space of $\vmu$. 

{\bf\em Generalized Clausius inequality.}  A combination of Eqs. (\ref{phi2psi}) and (\ref{psi2phi}a) yields a Clausius' inequality like relation:
\begin{equation}
\label{C-inequality}
    \varphi(\vx') + \vx'\cdot\vy' -
    \psi(\vy') \ge 0.
\end{equation}
The thermodynamics equilibrium is between the observed mean value $\vx'$ and their conjugate ``force'' $\vy'$.  When the equality holds, there is a relation between $\vx'$ and $\vy'$ which should be identified as a ``the equation of state'', with $\nabla_{\vx}\varphi=-\vy'$ and $\nabla_{\vy}\psi=\vx'$. When the $\vx'\neq\nabla_{\vy}\psi(\vy)$, the difference $\vx'\cdot\vy'-\psi(\vy')$ can be interpreted as the nonequilibrium heat and $\varphi$ is again as the entropy; then the inequality in (\ref{C-inequality}) becomes the Clausius' inequality.

{\bf\em Generalized Gibbs-Duhem equation.}  The celebrated Gibbs-Duhem equation in classical thermodynamics is a consequence of the entropy being an Eulerian degree $1$ homogeneous function. For the $\Phi(\vnu)$ in (\ref{ent-func}), thus we have
\begin{eqnarray}
  &\displaystyle 
  \Phi(\vnu) = \sum_{i=0}^n \nu_i\left(\frac{\partial\Phi}{\partial\nu_i}\right),\ 
  \sum_{i=0}^n \nu_i\left(\frac{\partial^2\Phi}{\partial\nu_i\partial\nu_j}\right) = 0, 
\nonumber\\
  &\displaystyle 
  \sum_{j=0}^n\rd\nu_j \sum_{i=0}^n \nu_i\left(\frac{\partial^2\Phi}{\partial\nu_i\partial\nu_j}\right) = 
  \sum_{i=0}^n \nu_i \sum_{j=0}^n 
  \left(\frac{\partial\mu_i}{\partial\nu_j}\right)\rd\nu_j = 0, 
\nonumber\\
  &\displaystyle 
  \text{ that is, }
    \sum_{i=0}^n \nu_i 
     \rd\mu_i = 0,
     \label{g-GDE}
\end{eqnarray} 
in which we have used (\ref{fe-func}b).  We identify (\ref{g-GDE}) as a generalized Gibbs-Duhem equation.

{\bf\em Discussion.}  Mathematical theory of probability deals with a set of elementary events $\mathcal{S}$, on which the probability $\vp$, and random variables $g$'s are introduced.  Applying this mathematics to real world, each ergodic dynamical system with state space $\mathcal{S}$ has its own unique steady-state probability distribution which can be obtained as the $\hat{\vnu}$ from i.i.d. sampling {\em ad infinitum}.  The entropy function in (\ref{ent-func}) arises in this context as a measure of the quantitative relationship between the assumed, ``hypothesis'' ($\vp$) and the observed ``data'' ($\vnu$ and $\hat{\vnu}$); as ``missing information'' or ``surprise'' \cite{bennaim-book}.  

Motivated by the analogy to Fourier analysis, our generalized Gibbs' theory seems to suggest that the notion of thermo-energetics is a powerful mathematical transformation of the statistical description; $\hat{\vnu}$ and $\vmu$ are simply two {\em representations} of a same physical reality; the former being statistical while the latter thermo-energetic.  With a fixed $\vp$, the theory of probability \cite{dembo-book} revealed a powerful, dual energetic representation for various different systems in terms of internal energy functions $\vmu$ \cite{qian_jctc}.  This fundamental dual between counting frequency and internal energy of course has been recognized by L. Boltzmann already in 1880s, when he was developing the statistical mechanics as a foundation of classical thermodynamics under the principle of {\em equal probability a priori}.  The present work shows that while the {\em probability and statistics} are fundamental as the foundation of thermodynamics, mechanics is not necessary.  A similar conclusion was reached in the 1925 thesis of L. Szilard \cite{szilard,mandelbrot_1964}.  

For sufficiently large $N$, the probability of observing a particular $\hat{\vnu}$ is asymptotically zero except $\hat{\vnu}=\vp$.  The significance of $\Phi\big(\hat{\vnu};\vp\big)$ is to provide a ``high-resolution magnifying glass'' for the asymptotically small 
\begin{equation} 
      \exp\Big\{-\hspace{-2pt}N\Phi\big(\hat{\vnu};\vp\big)\Big\}.
\label{equation17}      
\end{equation} 
This is known as the large deviations rate function in the modern theory of probability \cite{dembo-book}.  The entropy $-\Phi(\hat{\vnu},\vp)$ is a function of both $\hat{\vnu}$ and $\vp$; $\Phi(\hat{\vnu};\vp)\ge 0$ and $\Phi(\vp;\vp)=0$.  For a given $\vp$, it views each possible $\hat{\vnu}$ from a real system as a part of an entire class of systems under a common $\vp$, a {\em metastatistics}.  If one chooses the true steady state probability $\vpi$ of a particular system to replace $\vp$, then Eq. \ref{equation17} gives the probability distribution of the uncertainties in the measurement $\hat{\vnu}$ from $N$ samples.  The second-order Taylor expansion near $\vpi$, 
\[
  e^{-\hspace{-2pt}N\Phi(\hat{\vnu};\vpi)} 
   \simeq \exp\left[-N\sum_{i,j=0}^n 
   \frac{(\hat{\nu}_i-\pi_i)(\delta_{ij}-\pi_i)(\hat{\nu}_j-\pi_j)}{\pi_i}
   \right],
\]
is the central limit theorem for the statistics of counting frequency $\hat{\vnu}$, with $\text{Var}[\hat{\nu}_i]=\pi_i(1-\pi_i)/N$ and
$\text{Cov}[\hat{\nu}_i,\hat{\nu}_j]=-\pi_i\pi_j/N$.  This is not the fluctuations within the $\vpi$ of the system itself.  {\em Gibbs' theory of ensemble is about statistical measurements; not about a fluctuating system.}
 
We choose to present our theory with finite state space $\mathcal{S}$ for mathematical simplicity. Formal generalization to continuous state space is straight forward if mathematical rigor is not required.  Beyond the finite state space, it is well known that modern probability and the theory of measures encounter challenges, {\em c.f.}, de Finetti's treatment of infinite sets and the axiom of choice of nonempity subsets \cite{JaynesBook}.  

{\bf\em Acknowledgement.} I thank Jin Feng, Weishi Liu, Zhang-Ju Liu, Zhongmin Shen, Xiang Tang, and particularly Professor Jun Zhang, for many helpful discussions, and the support from Olga Jung Wan Endowed Professorship.

\appendix

\section{Appendix: Statistical counting {\em ad infinitum}
}

In this section, we provide the mathematical reasoning for stating ``entropy provides the infinitesimal probability for an observed frequency $\hat{\vnu}$ w.r.t. a probability prior $\vp$'', ``it characterizes the {\em relationship} between $\vnu$ and $\vp$ in a sampling process'', and the origin of Legendre-Fenchel transform in entropy analysis.  The counting of independent and identically distributed samples with state space $\mathcal{S}=\{0,\cdots,n\}$ yields $\vnu=(\nu_0,\cdots,\nu_n)$, a $(n+1)$-tuple of non-negative integers.  We call all the $\vnu$ with $N=\nu_0+\cdots+\nu_n$ a {\em simplex for counting}.  The simplex for counting grows with $N$, which we shall identify as ``time''.  With a given prior probability $\vp=(p_0,\cdots,p_n)$ on $\mathcal{S}$, statistical counting is a Markov process on a growing simplex, with probability:
\begin{equation}
\label{e1}
    P^{(N+1)}\big(\vnu\big) = 
    \sum_{k=0}^n p_k P^{(N)}\big(\vnu-\vdelta_k \big),
\end{equation}
in which $\vdelta_k = \big(0,\cdots,1,\cdots,0)$ is the unit vector for the $k^{th}$ component.  One can easily verify that 
\[
   P^{(N)}(\vnu) = \frac{N!}{\nu_0!\cdots\nu_n!} 
   p_0^{\nu_0}\cdots p_n^{\nu_n}
\] 
is a solution to (\ref{e1}).

One is interested in the limit of counting {\em ad infinitum}, when all the $\nu_i$'s are expected to tend infinity as $N\to\infty$.  On the increasing simplex for $\vnu$, the probability $P^{(N)}(\vnu)\to 0$.  However the properly normalized $\hat{\vnu}=\vnu/N$ converges, and $P^{(N)}$ as a function of the $\hat{\vnu}$ becomes sharper and sharper, concentrated around $\hat{\vnu}^*=\vp$.  To more precisely characterize this limiting situation, one introduces {\em counting frequency} $\hat{\vnu}=\vnu/N$. The space of $\hat{\vnu}$'s then is called a {\em probability simplex} $\mathscr{M}_n$; Eq. (\ref{e1}) then becomes
\begin{equation}
    \tilde{P}^{(N+1)}\big(\hat{\vnu}\big) =
    \sum_{k=0}^n p_k  \tilde{P}^{(N)}\Big( \big\{\tfrac{N+1}{N}\hat{\nu}_i-\tfrac{1}{N}\delta_{ik}\big\} \Big).
\end{equation} 
Its limit is a Dirac-$\delta$ function: $\tilde{P}^{(\infty)}=0$ for all $\hat{\vnu}\neq\vp$, and $\tilde{P}^{(\infty)}=\infty$ at $\hat{\vnu}=\vp$. However, ``a higher order'' infinitesimal analysis shows that \cite{dembo-book}:
\begin{equation}
    \lim_{N\to\infty}\frac{1}{N}
    \ln\tilde{P}^{(N)}\big(\hat{\vnu}\big) 
    = -\sum_{i=0}^n \hat{\nu}_i
    \ln\left(\frac{\hat{\nu}_i}{p_i}\right) = -\Phi\big(\hat{\vnu}\big).
\label{e-func}
\end{equation}
It is clear that entropy function $-\Phi(\vnu)$ represents the infinitesimal prior probability $e^{-N\Phi(\vnu)}$ on $\mathscr{M}_n$.  For two $\vnu$'s with different entropy values, $\Phi(\vnu)$ and $\Phi(\vnu')$, their probabilities $P^{\infty}(\vnu)/P^{\infty}(\vnu')=0$ if $\Phi(\vnu)>\Phi(\vnu')$.  This is the origin of the maximum entropy principle (MEP). 

To understand the limit $P^{(N)}(\vnu)\to 0$, one can also introduce the probability generating function  \cite{dembo-book}:
\begin{equation}
\label{gf}
      W^{(N)}\big(\vmu\big) = \sum_{\vnu}
    P^{(N)}\big(\vnu\big)
         e^{-\vmu\cdot\vnu},
\end{equation}
in which $\vmu\cdot\vnu=\mu_0\nu_0+\cdots+\mu_n\nu_n$.  Then Eq. (\ref{e1}) becomes
\begin{eqnarray} 
  && W^{(N+1)}\big(\vmu\big) = \sum_{\vnu}
  P^{(N+1)}\big(\vnu\big)e^{-\vmu\cdot\vnu} 
\nonumber\\  
  &=& \sum_{k=0}^n p_k e^{-\vmu\cdot\vdelta_k} \sum_{\vnu} P^{(N)}\big(\vnu-\vdelta_k \big)
  e^{-\vmu\cdot(\vnu-\vdelta_k) } 
\nonumber\\
  &=& W^{(N)}\big( 
     \vmu\big) e^{-\Psi(\vmu)},
\nonumber\\
 && \text{where }  
     \Psi(\vmu)=-\ln \sum_{k=0}^n 
     p_ke^{-\mu_k}.
\end{eqnarray} 
The free energy function $\Psi$ is meaningful for all finite $N$. This is why the partition function is valid even for small systems in Gibbs' theory of ensembles \cite{lu-qian-22}.  The Legendre-Fenchel transform of $\Psi(\vmu)$ is precisely the the right-hand-side of (\ref{e-func}):
\begin{eqnarray} 
  && \inf_{\vmu}\Big\{ 
  \vmu\cdot\vnu - \Psi\big(\vmu\big) \Big\}
\nonumber\\
&=& \inf_{\vmu} \left\{ 
  \sum_{i=0}^n\nu_i \ln e^{-\mu_i}
   +\ln\sum_{k=0}^n p_ke^{-\mu_k} 
  \right\} 
\\
    &=&  \inf_{\vmu} \left\{ 
  -\sum_{i=0}^n\nu_i \ln \left[ \frac{ e^{-\mu_i} }{ 
   \sum_{k=0}^n p_ke^{-\mu_k} 
   }\right] 
  \right\} =
-\sum_{i=0}^n \nu_i\ln\frac{\hat{\nu}_i}{
  p_i}, 
\nonumber 
\end{eqnarray} 
in which the optimal $e^{-\mu_i}\propto \nu_i/p_i$.  Legendre-Frenchel transform arises in the limit of $N\to\infty$ through the Laplace's method of evaluating asymptotic integrals, or the related Darwin-Fowler method of maximum term.  

\raggedright
\bibliography{main}

\end{document}